\documentclass[12pt]{article}

\begin{document}

\title{Horizons and Geodesics of Black Ellipsoids }
\author{Sergiu I. Vacaru \thanks{%
E-mail address:\ vacaru@fisica.ist.utl.pt, ~~ sergiu$_{-}$vacaru@yahoo.com,\
} \\
{\small \textit{Centro Multidisciplinar de Astrofisica - CENTRA,
Departamento de Fisica,}}\\
{\small \textit{Instituto Superior Tecnico, Av. Rovisco Pais 1, Lisboa,
1049-001, Portugal}}\\
}
\date{July 21, 2003}
\maketitle

\begin{abstract}
We analyze the horizon and geodesic structure of a class of 4D off--diagonal
metrics with deformed spherical symmetries, which are exact solutions of the
vacuum Einstein equations with anholonomic variables. The maximal analytic
extension of the ellipsoid type metrics are constructed and the Penrose
diagrams are analyzed with respect to adapted frames. We prove that for
small deformations (small eccentricities) there are such metrics that the
geodesic behaviour is similar to the Schwarzcshild one. We conclude that
some vacuum static and stationary ellipsoid configurations \cite{v,v1} may
describe black ellipsoid objects.

\vskip5pt.

Pacs 04.20.Jb,\ 04.70.-s,\ 04.70.Bw

\end{abstract}



\section{Introduction}

Recently, the off--diagonal metrics were considered in a new manner by
diagonalizing them with respect to anholonomic frames with associated
nonlinear connection structure \cite{v,v1,vsbd}. There were constructed new
classes of exact solutions of Einstein's equations in three (3D), four (4D)
and five (5D) dimensions. Such vacuum solutions posses a generic geometric
local anisotropy (\textit{e.g.} static black hole and cosmological solutions
with ellipsoidal or toroidal symmetry, various soliton--dilaton 2D and 3D
configurations in 4D gravity, and wormholes and flux tubes with anisotropic
polarizations and/or running constants with different extensions to
backgrounds of rotation ellipsoids, elliptic cylinders, bipolar and toroidal
symmetry and anisotropy).

A number of off--diagonal metrics were investigated in higher dimensional
gravity (see, for instance, the Salam, Strathee, Percacci and
Randjbar--Daemi works \cite{sal} which showed that including off--diagonal
components in higher dimensional metrics is equi\-valent to including $%
U(1),SU(2)$ and $SU(3)$ gauge fields. There are various generalizations of
the Kaluza--Klein gravity when the off--diagonal metrics and their
compactifications are considered in order to reduce the vacuum 5D gravity to
effective Einstein gravity and Abelian or non--Abelian gauge theories. One
has also constructed 4D exact solutions of Einstein equations with matter
fields and cosmological constants like black torus and black strings induced
from some 3D black hole configurations by considering 4D off--diagonal
metrics whose curvature scalar splits equivalently into a curvature term for
a diagonal metric together with a cosmological constant term and/or a
Lagrangian for gauge (electromagnetic) field \cite{lemos}.

For some particular off--diagonal metric ansatz and redefinitions of
Lagrangians we can model certain effective (diagonal metric) gravitational
and matter fields interactions. However, in general, the vacuum
gravitational dynamics can not be associated to any matter field
contributions. Our aim is to investigate such off--diagonal vacuum
gravitational configurations (defined by anholonomic frames with associated
nonlinear connection structure) which describe black hole solutions with
deformed horizons, for instance, with a static ellipsoid hypersurface.

In this paper we construct the maximal analytic extension of a class of
static metrics with deformed spherical symmetry (containing as particular
cases ellipsoid configurations). We analyze the Penrose diagrams and compare
the results with those for the Reissner--Nordstrom solution. Then we state
the conditions when the geodesic congruence with 'ellipsoid' type symmetry
can be reduced to the Schwarzschild configuration. We argue that in this
case we may generate some static black ellipsoid solutions which, for
corresponding parametrizations of off--diagonal metric coefficients, far
away from the horizon, satisfy the asymptotic conditions of the Minkowski
space--time.

The paper has the following structure: In Sec. 2 we present the necessary
formulas on off--diagonal metrics and anholonomic frames with associated
nonlinear connection structure and write the vacuum Einstein equations with
anholonomic variables corresponding to a general off--diagonal metric
ansatz. In Sec. 3 we define a class of static anholonomic deformations of
the Scwarzschild metric to some off--diagonal metrics having their
coefficients very similar to the Reissner--N\"{o}rdstrom but written with
respect to adapted frames and defined as vacuum configurations. In Sec. 4,
for static small ellipsoid deformations, we construct the maximal analytic
extension of such metrics and analyze their horizon structure. Section 5
contains a study of the conditions when the geodesic behaviour of
ellipsoidal metrics can be congruent to the Schwarzschild one. Conclusions
are given in Sec. 6.

\section{Off--diagonal Metrics and \newline
Anholonomic Frames}

Let $V^{3+1}$be a 4D pseudo--Riemannian space--time enabled with local
coordinates $u^{\alpha }=\left( x^{i},y^{a}\right) $ where the indices of
type $i,j,k,...$ run values $1$ and $2$ and the indices $a,b,c,...$ take
values $3$ and $4;$ $\ y^{3}=v=\varphi $ and $y^{4}=t$ are considered
respectively as the ''anisotropic'' and time like coordinates (subjected to
some constraints). This space--time $V^{3+1}$ we may be provided with a
general anholonomic frame structure of tetrads, or vierbiends,
\begin{equation}
e_{\alpha }=A_{\alpha }^{\beta }\left( u^{\gamma }\right) \partial /\partial
u^{\beta },  \label{transftet}
\end{equation}%
satisfying some anholonomy \ relations
\begin{equation}
e_{\alpha }e_{\beta }-e_{\beta }e_{\alpha }=W_{\alpha \beta }^{\gamma
}\left( u^{\varepsilon }\right) e_{\gamma },  \label{anhol}
\end{equation}%
where $W_{\alpha \beta }^{\gamma }\left( u^{\varepsilon }\right) $ are
called the coefficients of anholonomy. One defines a 'holonomic' coordinate
frames, for instance, a coordinate frame, $e_{\alpha }=\partial _{\alpha
}=\partial /\partial u^{\alpha },$ as a set of tetrads which satisfy the
holonomy conditions
\[
\partial _{\alpha }\partial _{\beta }-\partial _{\beta }\partial _{\alpha
}=0.
\]

The 4D line element
\begin{equation}
ds^{2}={g}_{\alpha \beta }\left( x^{i},v\right) du^{\alpha }du^{\beta },
\label{cmetric4}
\end{equation}%
is parametrized by an ansatz {%
\begin{equation}
{g}_{\alpha \beta }=\left[
\begin{array}{cccc}
g_{1}+w_{1}^{\ 2}h_{3}+n_{1}^{\ 2}h_{4} & w_{1}w_{2}h_{3}+n_{1}n_{2}h_{4} &
w_{1}h_{3} & n_{1}h_{4} \\
w_{1}w_{2}h_{3}+n_{1}n_{2}h_{4} & g_{2}+w_{2}^{\ 2}h_{3}+n_{2}^{\ 2}h_{4} &
w_{2}h_{3} & n_{2}h_{4} \\
w_{1}h_{3} & w_{2}h_{3} & h_{3} & 0 \\
n_{1}h_{4} & n_{2}h_{4} & 0 & h_{4}%
\end{array}%
\right] ,  \label{ansatzc4}
\end{equation}%
} with $g_{i}=g_{i}\left( x^{i}\right) ,h_{a}=h_{ai}\left( x^{k},v\right) $
and $n_{i}=n_{i}\left( x^{k},v\right) $ \ being some functions of necessary
smoothly class or even sigular in some points and finite regions. The
coefficients $g_{i}$ depend only on ''holonomic'' variables $x^{i}$ but the
rest of coefficients may also depend on one ''anisotropic'' (anholonomic)
variable $y^{3}=v;$ the ansatz does not depend on the time variable $%
y^{4}=t: $ we shall search for static solutions.

We can diagonalize the line element (\ref{cmetric4}),
\begin{equation}
\delta s^{2}=g_{1}(dx^{1})^{2}+g_{2}(dx^{2})^{2}+h_{3}(\delta
v)^{2}+h_{4}(\delta y^{4})^{2},  \label{dmetric4}
\end{equation}%
with respect to the anholonomic co--frame
\begin{equation}
\delta ^{\alpha }=(d^{i}=dx^{i},\delta ^{a}=dy^{a}+N_{i}^{a}dx^{i})=\left(
d^{i},\delta v=dv+w_{i}dx^{i},\delta y^{4}=dy^{4}+n_{i}dx^{i}\right)
\label{ddif4}
\end{equation}%
which is dual to the anholonomic frame
\begin{equation}
\delta _{\alpha }=(\delta _{i}=\partial _{i}-N_{i}^{a}\partial _{a},\partial
_{b})=\left( \delta _{i}=\partial _{i}-w_{i}\partial _{3}-n_{i}\partial
_{4},\partial _{3},\partial _{4}\right) ,  \label{dder4}
\end{equation}%
where $\partial _{i}=\partial /\partial x^{i}$ and $\partial _{b}=\partial
/\partial y^{b}$ are usual partial derivatives. The tetrads $\delta _{\alpha
}$ and $\delta ^{\alpha }$ are anholonomic because, in general, they satisfy
the anholonomy relations (\ref{anhol}) with some non--trivial coefficients,
\[
W_{ij}^{a}=\delta _{i}N_{j}^{a}-\delta
_{j}N_{i}^{a},~W_{ia}^{b}=-~W_{ai}^{b}=\partial _{a}N_{i}^{b}.
\]%
The anholonomy is induced by the coefficients $N_{i}^{3}=w_{i}$ and $%
N_{i}^{4}=n_{i}$ which ''elongate'' partial derivatives and differentials if
we are working with respect to anholonomic frames. This results in a more
sophisticate differential and integral calculus (as in the tetradic and/or
spinor gravity), but simplifies substantially the tensor computations,
because we are dealing with diagonalized metrics.

With respect to the anholonomic frames (\ref{dder4}) and (\ref{ddif4}),
there is a linear connection, called the canonical distinguished linear
connection, which is similar to the metric connection introduced by the
Christoffel symbols in the case of holonomic bases, i. e. being constructed
only from the metric components and satisfying the metricity conditions $%
D_{\alpha }g_{\beta \gamma }=0.$ It is parametrized by the coefficients,\ $%
\Gamma _{\ \beta \gamma }^{\alpha }=\left( L_{\ jk}^{i},L_{\ bk}^{a},C_{\
jc}^{i},C_{\ bc}^{a}\right) $ with the coefficients
\begin{eqnarray}
L_{\ jk}^{i} &=&\frac{1}{2}g^{in}\left( \delta _{k}g_{nj}+\delta
_{j}g_{nk}-\delta _{n}g_{jk}\right) ,  \label{dcon} \\
L_{\ bk}^{a} &=&\partial _{b}N_{k}^{a}+\frac{1}{2}h^{ac}\left( \delta
_{k}h_{bc}-h_{dc}\partial _{b}N_{k}^{d}-h_{db}\partial _{c}N_{k}^{d}\right) ,
\nonumber \\
C_{\ jc}^{i} &=&\frac{1}{2}g^{ik}\partial _{c}g_{jk},\ C_{\ bc}^{a}=\frac{1}{%
2}h^{ad}\left( \partial _{c}h_{db}+\partial _{b}h_{dc}-\partial
_{d}h_{bc}\right) ,  \nonumber
\end{eqnarray}%
computed for the ansatz (\ref{ansatzc4}), which induce a linear covariant
derivative locally adapted to the nonlinear connection structure
(N--connection, see details, for instance, in Refs. \cite{ma,v,vth}). We
note that on (pseudo) Riemannian spaces the N--connection is an object
completely defined by anholonomic frames, when the coefficients of tetradic
transform (\ref{transftet}), $A_{\alpha }^{\beta }\left( u^{\gamma }\right)
, $ are parametrized explicitly via certain values $\left( N_{i}^{a},\delta
_{i}^{j},\delta _{b}^{a}\right) ,$ where $\delta _{i}^{j}$ $\ $and $\delta
_{b}^{a}$ are the Kronecker symbols. By straightforward calculations we can
compute (see, for instance Ref. \cite{mtw}) that the coefficients of the
Levi Civita metric connection
\[
\Gamma _{\alpha \beta \gamma }^{\nabla }=g\left( \delta _{\alpha },\nabla
_{\gamma }\delta _{\beta }\right) =g_{\alpha \tau }\Gamma _{\beta \gamma
}^{\nabla \tau },\,
\]%
associated to a covariant derivative operator $\nabla ,$ satisfying the
metricity condition $\nabla _{\gamma }g_{\alpha \beta }=0$ for $g_{\alpha
\beta }=\left( g_{ij},h_{ab}\right) ,$
\begin{equation}
\Gamma _{\alpha \beta \gamma }^{\nabla }=\frac{1}{2}\left[ \delta _{\beta
}g_{\alpha \gamma }+\delta _{\gamma }g_{\beta \alpha }-\delta _{\alpha
}g_{\gamma \beta }+g_{\alpha \tau }W_{\gamma \beta }^{\tau }+g_{\beta \tau
}W_{\alpha \gamma }^{\tau }-g_{\gamma \tau }W_{\beta \alpha }^{\tau }\right]
,  \label{lcsym}
\end{equation}%
are given with respect to the anholonomic basis (\ref{ddif4}) by the
coefficients
\begin{equation}
\Gamma _{\beta \gamma }^{\nabla \tau }=\left( L_{\ jk}^{i},L_{\ bk}^{a},C_{\
jc}^{i}+\frac{1}{2}g^{ik}\Omega _{jk}^{a}h_{ca},C_{\ bc}^{a}\right) ,
\label{lccon}
\end{equation}%
where
\[
\Omega _{jk}^{a}=\delta _{k}N_{j}^{a}-\delta _{j}N_{k}^{a}.
\]%
We emphasize that the (pseudo) Riemannian space-times admit non--trivial
torsion components,
\begin{equation}
T_{ja}^{i}=-T_{aj}^{i}=C_{ja}^{i},T_{jk}^{a}=-T_{kj}^{a}=\Omega
_{kj}^{a},T_{bk}^{a}=-T_{kb}^{a}=\partial _{b}N_{k}^{a}-L_{bk}^{a},
\label{torsion}
\end{equation}%
if off--diagonal metrics and anholomomic frames are introduced into
consideration. This is a ''pure'' anholonomic frame effect: the torsion
vanishes for the Levi Civita connection stated with respect to a coordinate
frame, but even this metric connection contains some torsion coefficients if
it is defined with respect to anholonomic frames (this follows from the $W$%
--terms in (\ref{lcsym})). We can conclude that the Einstein theory
transforms into an effective Einstein--Cartan theory with anholonomically
induced torsion if the general relativity is formulated with respect to
general frame bases (both holonomic and anholonomic).

Another specific property of off--diagonal (pseudo) Riemannian metrics is
that they can define different classes of linear connections which satisfy
the metricity conditions for a given metric, or inversely, there is a
certain class of metrics which satisfy the metricity conditions for a given
linear connection. \ This result was originally obtained by A. Kawaguchi %
\cite{kaw} (Details can be found in Ref. \cite{ma}, see Theorems 5.4 and 5.5
in Chapter III, formulated for vector bundles; here we note that similar
proofs hold also on manifolds enabled with anholonomic frames associated to
a N--connection structure.)

With respect to anholonomic frames, we can not distinguish the Levi Civita
connection as the unique both metric and torsionless one. For instance, both
linear connections (\ref{dcon}) and (\ref{lccon}) contain anholonomically
induced torsion coefficients, are compatible with the same metric and
transform into the usual Levi Civita coefficients for vanishing
N--connection and ''pure'' holonomic coordinates. This means that to an
off--diagonal metric in general relativity one may be associated different
covariant differential calculi, all being compatible with the same metric
structure (like in the non--commutative geometry, which is not a surprising
\ fact because the anolonomic frames satisfy by definition some
non--commutative relations (\ref{anhol})). In such cases we have to select a
particular type of connection following some physical or geometrical
arguments, or to impose some conditions when there is a single compatible
linear connection constructed only from the metric and N--coefficients. We
note that if $\Omega _{jk}^{a}=0$ the connections (\ref{dcon}) and (\ref%
{lccon}) coincide, i. e. $\Gamma _{\ \beta \gamma }^{\alpha }=\Gamma _{\beta
\gamma }^{\nabla \alpha }.$

The vacuum Einstein equations $R_{\alpha }^{\beta }=0$ computed for the
metric (\ref{dmetric4}) and connection (\ref{dcon}) with respect to
anholonomic frames (\ref{ddif4}) and (\ref{dder4}) transform into a system
of partial differential equations with anholonomic variables \cite{v,v1,vth}%
,
\begin{eqnarray}
R_{1}^{1}=R_{2}^{2}=-\frac{1}{2g_{1}g_{2}}[g_{2}^{\bullet \bullet }-\frac{%
g_{1}^{\bullet }g_{2}^{\bullet }}{2g_{1}}-\frac{(g_{2}^{\bullet })^{2}}{%
2g_{2}}+g_{1}^{^{\prime \prime }}-\frac{g_{1}^{^{\prime }}g_{2}^{^{\prime }}%
}{2g_{2}}-\frac{(g_{1}^{^{\prime }})^{2}}{2g_{1}}] &=&0,  \label{ricci1a} \\
R_{3}^{3}=R_{4}^{4}=-\frac{\beta }{2h_{3}h_{4}}=-\frac{1}{2h_{3}h_{4}}\left[
h_{4}^{\ast \ast }-h_{4}^{\ast }\left( \ln \sqrt{|h_{3}h_{4}|}\right) ^{\ast
}\right] &=&0,  \label{ricci2a} \\
R_{3i}=-w_{i}\frac{\beta }{2h_{4}}-\frac{\alpha _{i}}{2h_{4}} &=&0,
\label{ricci3a} \\
R_{4i}=-\frac{h_{4}}{2h_{3}}\left[ n_{i}^{\ast \ast }+\gamma n_{i}^{\ast }%
\right] &=&0,  \label{ricci4a}
\end{eqnarray}%
where
\begin{equation}
\alpha _{i}=\partial _{i}h_{4}^{\ast }-h_{4}^{\ast }\partial _{i}\ln \sqrt{%
|h_{3}h_{4}|},~\gamma =3h_{4}^{\ast }/2h_{4}-h_{3}^{\ast }/h_{3},
\label{abc}
\end{equation}%
and the partial derivatives are denoted $g_{1}^{\bullet }=\partial
g_{1}/\partial x^{1},g_{1}^{^{\prime }}=\partial g_{1}/\partial x^{2}$ and $%
h_{3}^{\ast }=\partial h_{3}/\partial v.$ We additionally impose the
condition $\delta _{i}N_{j}^{a}=\delta _{j}N_{i}^{a}$ in order to have $%
\Omega _{jk}^{a}=0$ which may be satisfied, for instance, if
\[
w_{1}=w_{1}\left( x^{1},v\right) ,n_{1}=n_{1}\left( x^{1},v\right)
,w_{2}=n_{2}=0,
\]%
or, inversely, if
\[
w_{1}=n_{1}=0,w_{2}=w_{2}\left( x^{2},v\right) ,n_{2}=n_{2}\left(
x^{2},v\right) .
\]%
In this case the canonical connection (\ref{dcon}) is equivalent to the Levi
Civita connection (\ref{lccon}) written with respect to anholonomic frames
and containing some non--trivial coefficients of induced torsion (\ref%
{torsion}).

In this paper we shall select a class of static solutions parametrized by
the conditions
\begin{equation}
w_{1}=w_{2}=n_{2}=0.  \label{cond1}
\end{equation}

\ The system of equations (\ref{ricci1a})--(\ref{ricci4a}) can be integrated
in general form \cite{vth}. Physical solutions are selected following some
additional boundary conditions, imposed types of symmetries, nonlinearities
and singular behaviour and compatibility in the locally anisotropic limits
with some well known exact solutions.

\section{Anholonomic Deformations of \newline
the Schwarzschild Solution}

As a background for our investigations we consider an off--diagonal metric
ansatz%
\begin{eqnarray}
\delta s^{2} &=&[-\left( 1-\frac{2m}{r}+\frac{\varepsilon }{r^{2}}\right)
^{-1}dr^{2}-r^{2}q(r)d\theta ^{2}  \label{sch} \\
&&-\eta _{3}\left( r,\varphi \right) r^{2}\sin ^{2}\theta d\varphi ^{2}+\eta
_{4}\left( r,\varphi \right) \left( 1-\frac{2m}{r}+\frac{\varepsilon }{r^{2}}%
\right) \delta t^{2}]  \nonumber
\end{eqnarray}%
where the ''polarization'' functions $\eta _{3,4}$ are decomposed on a small
parameter $\varepsilon ,0<\varepsilon \ll 1,$
\begin{eqnarray}
\eta _{3}\left( r,\varphi \right)  &=&\eta _{3[0]}\left( r,\varphi \right)
+\varepsilon \lambda _{3}\left( r,\varphi \right) +\varepsilon ^{2}\gamma
_{3}\left( r,\varphi \right) +...,  \label{decom1} \\
\eta _{4}\left( r,\varphi \right)  &=&1+\varepsilon \lambda _{4}\left(
r,\varphi \right) +\varepsilon ^{2}\gamma _{4}\left( r,\varphi \right) +...,
\nonumber
\end{eqnarray}%
and
\[
\delta t=dt+n_{1}\left( r,\varphi \right) dr
\]%
for $n_{1}\sim \varepsilon ...+\varepsilon ^{2}$ terms. The functions $\eta
_{3,4}\left( r,\varphi \right) $ and $n_{1}\left( r,\varphi \right) $ will
be found as the metric will define a solution of the vacuum Einstein
equations generated by small deformations of the spherical static symmetry
on a small positive parameter $\varepsilon $ (in the limits $\varepsilon
\rightarrow 0$ and $q,\eta _{3,4}\rightarrow 1$ we have just the
Schwarzschild solution for a point particle of mass $m).$ The metric (\ref%
{sch}) is a particular case of a class of exact solutions constructed in %
\cite{v,v1,vth}.

The condition of vanishing of the metric coefficient before $\delta t^{2}$%
\begin{eqnarray}
\eta _{4}\left( r,\varphi \right) \left( 1-\frac{2m}{r}+\frac{\varepsilon }{%
r^{2}}\right)  &=&1-\frac{2m}{r}+\varepsilon \frac{\Phi _{4}}{r^{2}}%
+\varepsilon ^{2}\Theta _{4}=0,  \label{hor1} \\
\Phi _{4} &=&\lambda _{4}\left( r^{2}-2mr\right) +1  \nonumber \\
\Theta _{4} &=&\gamma _{4}\left( 1-\frac{2m}{r}\right) +\lambda _{4},
\nonumber
\end{eqnarray}%
defines a rotation ellipsoid configuration if
\[
\lambda _{4}=\left( 1-\frac{2m}{r}\right) ^{-1}(\cos \varphi -\frac{1}{r^{2}}%
)
\]%
and
\[
\gamma _{4}=-\lambda _{4}\left( 1-\frac{2m}{r}\right) ^{-1}.
\]%
In the first order on $\varepsilon $ one obtains \ a zero value for the
coefficient before $\delta t^{2}$ if
\begin{equation}
r_{+}=\frac{2m}{1+\varepsilon \cos \varphi }=2m[1-\varepsilon \cos \varphi ],
\label{ebh}
\end{equation}%
which is the equation for a 3D ellipsoid like hypersurface with a small
eccentricity $\varepsilon .$ In general, we can consider arbitrary pairs of
functions $\lambda _{4}(r,\theta ,\varphi )$ and $\gamma _{4}(r,\theta
,\varphi )$ (for $\varphi $--anisotropies, \ $\lambda _{4}(r,\varphi )$ and $%
\gamma _{4}(r,\varphi ))$ which may be singular for some values of $r,$ or
on some hypersurvaces $r=r\left( \theta ,\varphi \right) $ ($r=r(\varphi )).$

The simplest way to analyze the condition of vanishing of the metric
coefficient before $\delta t^{2}$ is to choose $\gamma _{4}$ and $\lambda
_{4}$ as to have $\Theta =0.$ In this case we find from \ (\ref{hor1})%
\begin{equation}
r_{\pm }=m\pm m\sqrt{1-\varepsilon \frac{\Phi }{m^{2}}}=m\left[ 1\pm \left(
1-\varepsilon \frac{\Phi _{4}}{2m^{2}}\right) \right]  \label{hor1a}
\end{equation}%
where $\Phi _{4}\left( r,\varphi \right) $ is taken for $r=2m.$

Having introduced a new radial coordinate
\begin{equation}
\xi =\int dr\sqrt{|1-\frac{2m}{r}+\frac{\varepsilon }{r^{2}}|}  \label{int2}
\end{equation}%
and defined
\begin{equation}
h_{3}=-\eta _{3}(\xi ,\varphi )r^{2}(\xi )\sin ^{2}\theta ,\ h_{4}=1-\frac{2m%
}{r}+\varepsilon \frac{\Phi _{4}}{r^{2}},  \label{sch1q}
\end{equation}%
for $r=r\left( \xi \right) $ found as the inverse function after integration
in (\ref{int2}), we write the metric (\ref{sch}) as
\begin{eqnarray}
ds^{2} &=&-d\xi ^{2}-r^{2}\left( \xi \right) q\left( \xi \right) d\theta
^{2}+h_{3}\left( \xi ,\theta ,\varphi \right) \delta \varphi
^{2}+h_{4}\left( \xi ,\theta ,\varphi \right) \delta t^{2},  \label{sch1} \\
\delta t &=&dt+n_{1}\left( \xi ,\varphi \right) d\xi ,  \nonumber
\end{eqnarray}%
where the coefficient $n_{1}$ is taken to solve the equation (\ref{ricci4a})
and to satisfy the (\ref{cond1}). $\ $

Let us define the conditions when the coefficients of metric (\ref{sch})
define solutions of the vacuum Einstein equations: For $%
g_{1}=-1,g_{2}=-r^{2}\left( \xi \right) q\left( \xi \right) $ and arbitrary $%
h_{3}(\xi ,\theta ,\varphi )$ and $h_{4}\left( \xi ,\theta \right) $ we get
solutions the equations (\ref{ricci1a})--(\ref{ricci3a}). If $h_{4}$ depends
on anisotropic variable $\varphi ,$ the equation (\ref{ricci2a}) may be
solved if
\begin{equation}
\sqrt{|\eta _{3}|}=\eta _{0}\left( \sqrt{|\eta _{4}|}\right) ^{\ast }
\label{conda}
\end{equation}%
for $\eta _{0}=const.$ Considering decompositions of type (\ref{decom1}) we
put $\eta _{0}=\eta /\varepsilon $ where the constant $\eta $ is taken as to
have $\sqrt{|\eta _{3}|}=1$ in the limits
\begin{equation}
\frac{\left( \sqrt{|\eta _{4}|}\right) ^{\ast }\rightarrow 0}{\varepsilon
\rightarrow 0}\rightarrow \frac{1}{\eta }=const.  \label{condb}
\end{equation}%
These conditions are satisfied if the functions $\eta _{3[0]},$ $\lambda
_{3,4}$ and $\gamma _{3,4}$ are related via relations
\[
\sqrt{|\eta _{3[0]}|}=\frac{\eta }{2}\lambda _{4}^{\ast },\lambda _{3}=\eta
\sqrt{|\eta _{3[0]}|}\gamma _{4}^{\ast }
\]%
for arbitrary $\gamma _{3}\left( r,\varphi \right) .$ In this paper we
select only such solutions which satisfy the conditions (\ref{conda}) and (%
\ref{condb}).

In order to consider linear infinitezimal extensions on $\varepsilon $ of
the Schwarzschild metric we may write the solution of (\ref{ricci4a}) as
\[
n_{1}=\varepsilon \widehat{n}_{1}\left( \xi ,\varphi \right)
\]%
where
\begin{eqnarray}
\widehat{n}_{1}\left( \xi ,\varphi \right) &=&n_{1[1]}\left( \xi \right)
+n_{1[2]}\left( \xi \right) \int d\varphi \ \eta _{3}\left( \xi ,\varphi
\right) /\left( \sqrt{|\eta _{4}\left( \xi ,\varphi \right) |}\right)
^{3},\eta _{4}^{\ast }\neq 0;  \label{auxf4} \\
&=&n_{1[1]}\left( \xi \right) +n_{1[2]}\left( \xi \right) \int d\varphi \
\eta _{3}\left( \xi ,\varphi \right) ,\eta _{4}^{\ast }=0;  \nonumber \\
&=&n_{1[1]}\left( \xi \right) +n_{1[2]}\left( \xi \right) \int d\varphi
/\left( \sqrt{|\eta _{4}\left( \xi ,\varphi \right) |}\right) ^{3},\eta
_{3}^{\ast }=0;  \nonumber
\end{eqnarray}%
with the functions $n_{k[1,2]}\left( \xi \right) $ to be stated by boundary
conditions.

The data
\begin{eqnarray}
g_{1} &=&-1,g_{2}=-r^{2}(\xi )q(\xi ),  \label{data} \\
h_{3} &=&-\eta _{3}(\xi ,\varphi )r^{2}(\xi )\sin ^{2}\theta ,~h_{4}=1-\frac{%
2m}{r}+\varepsilon \frac{\Phi _{4}}{r^{2}},  \nonumber \\
w_{1,2} &=&0,n_{1}=\varepsilon \widehat{n}_{1}\left( \xi ,\varphi \right)
,n_{2}=0,  \nonumber
\end{eqnarray}%
for the metric (\ref{sch}) define a class of solutions of the vacuum
Einstein equations with non--trivial polarization function $\eta _{3}$ and
extended on parameter $\varepsilon $ up to the second order (the
polarization functions being taken as to make zero the second order
coefficients). Such solutions are generated by small deformations (in
particular cases of rotation ellipsoid symmetry) of the Schwarschild metric.

We can relate our solutions with some small deformations of the
Schwar\-zschild metric, as well we can satisfy the asymptotically flat
condition, if we chose such functions $n_{k[1,2]}\left( x^{i}\right) $ as $%
n_{k}\rightarrow 0$ for $\varepsilon \rightarrow 0$ and $\eta
_{3}\rightarrow 1.$ These functions have to be selected as to vanish far
away from the horizon, for instance, like $\sim 1/r^{1+\tau },\tau >0,$ for
long distances $r\rightarrow \infty .$

\section{ Analytic Extensions of Ellipsoid Metrics}

The metric (\ref{sch}) (equivalently (\ref{sch1})) considered with respect
to the \ anholonomic frame (\ref{ddif4}) has a number of similarities with
the Schwrzschild and Reissner--N\"{o}rdstrom solutions. If we identify $%
\varepsilon $ with $e^{2},$ we get a static metric with effective
''electric'' charge induced by a small, quadratic on $\varepsilon ,$
off--diagonal metric extension. The coefficients of this metric are similar
to those from the Reissner--N\"{o}rdstrom solution but additionally to the
mentioned frame anholonomy there are additional polarizations by the
functions $h_{3[0]},\eta _{3,4}$ and $n_{1}.$ Another very important
property is that the deformed metric was stated to define a vacuum solution
of the Einstein equations which differs substantially from the usual
Reissner--N\"{o}rdstrom metric being an exact static solution of the
Einstein--Maxwell equations.

For $\varepsilon \rightarrow 0$ and $h_{3[0]}\rightarrow 1$ the metric (\ref%
{sch}) transforms into the usual Schwarzschild metric. A solution with
ellipsoid symmetry can be selected by a corresponding condition of vanishing
of the coefficient before $\delta t$ which defines an ellipsoidal
hypersurface like for the Kerr metric, but in our case the metric is
non--rotating.

The metric (\ref{sch}) has a singular behaviour for $r=r_{\pm },$ see (\ref%
{hor1a}). The aim of this section is to prove that this way we have
constructed a solution of the vacuum Einstein equations with an
''anisotropic'' horizon being a small deformation on parameter $\varepsilon $
of the Schwarzschild's solution horizon. We may analyze the anisotropic
horizon's properties for some fixed ''direction'' given in a smooth vecinity
of any values $\varphi =\varphi _{0}$ and $r_{+}=r_{+}\left( \varphi
_{0}\right) .$ $\ $The final conclusions will be some general ones for
arbitrary $\varphi $ when the explicit values of coefficients will have a
parametric dependence on angular coordinate $\varphi .$ The metrics (\ref%
{sch}) and (\ref{sch1}) are regular in the regions I ($\infty >r>r_{+}^{\Phi
}),$ II ($r_{+}^{\Phi }>r>r_{-}^{\Phi })$ and III$\;(r_{-}^{\Phi }>r>0).$ As
in the Schwarzschild, Reissner--N\"{o}rdstrom and Kerr cases these
singularities can be removed by introducing suitable coordinates and
extending the manifold to obtain a maximal analytic extension \cite%
{gb,carter}. We have similar regions as in the Reissner--N\"{o}rdstrom
space--time, but with just only one possibility $\varepsilon <1$ instead of
three relations for static electro--vacuum cases ($%
e^{2}<m^{2},e^{2}=m^{2},e^{2}>m^{2};$ where $e$ and $m$ are correspondingly
the electric charge and mass of the point particle in the Reissner--N\"{o}%
rdstrom metric). So, we may consider the usual Penrose's diagrams as for a
particular case of the Reissner--N\"{o}rdstrom space--time but keeping in
mind that such diagrams and horizons have an additional parametrization on
an angular coordinate.

To construct the maximally extended manifold, we proceed in steps analogous
to those in the Schwarzschild case (see details, for instance, in Ref. \cite%
{haw})). We introduce a new coordinate
\[
r^{\Vert }=\int dr\left( 1-\frac{2m}{r}+\frac{\varepsilon }{r^{2}}\right)
^{-1}
\]%
for $r>r_{+}^{1}$ and find explicitly the coodinate
\begin{equation}
r^{\Vert }=r+\frac{(r_{+}^{1})^{2}}{r_{+}^{1}-r_{-}^{1}}\ln (r-r_{+}^{1})-%
\frac{(r_{-}^{1})^{2}}{r_{+}^{1}-r_{-}^{1}}\ln (r-r_{-}^{1}),  \label{r1}
\end{equation}%
where $r_{\pm }^{1}=r_{\pm }^{\Phi }$ with $\Phi =1.$ If $r$ is expressed as
a function on $\xi ,$ than $r^{\Vert }$ can be also expressed as a function
on $\xi $ depending additionally on some parameters.

Defining the advanced and retarded coordinates, $v=t+r^{\Vert }$ and $%
w=t-r^{\Vert },$ with corresponding elongated differentials
\[
\delta v=\delta t+dr^{\Vert }\mbox{ and }\delta w=\delta t-dr^{\Vert }
\]%
the metric (\ref{sch1}) takes the form%
\[
\delta s^{2}=-r^{2}(\xi )q(\xi )d\theta ^{2}-\eta _{3}(\xi ,\varphi
_{0})r^{2}(\xi )\sin ^{2}\theta \delta \varphi ^{2}+(1-\frac{2m}{r(\xi )}%
+\varepsilon \frac{\Phi _{4}(r,\varphi _{0})}{r^{2}(\xi )})\delta v\delta w,
\]%
where (in general, in non--explicit form) $r(\xi )$ is a function of type $%
r(\xi )=r\left( r^{\Vert }\right) =$ $r\left( v,w\right) .$ Introducing new
coordinates $(v^{\prime \prime },w^{\prime \prime })$ by%
\[
v^{\prime \prime }=\arctan \left[ \exp \left( \frac{r_{+}^{1}-r_{-}^{1}}{%
4(r_{+}^{1})^{2}}v\right) \right] ,w^{\prime \prime }=\arctan \left[ -\exp
\left( \frac{-r_{+}^{1}+r_{-}^{1}}{4(r_{+}^{1})^{2}}w\right) \right]
\]%
and multiplying the last term on the conformal factor we obtain
\begin{equation}
\delta s^{2}=-r^{2}q(r)d\theta ^{2}-\eta _{3}(r,\varphi _{0})r^{2}\sin
^{2}\theta \delta \varphi ^{2}+64\frac{(r_{+}^{1})^{4}}{%
(r_{+}^{1}-r_{-}^{1})^{2}}(1-\frac{2m}{r(\xi )}+\varepsilon \frac{\Phi
_{4}(r,\varphi _{0})}{r^{2}(\xi )})\delta v^{\prime \prime }\delta w^{\prime
\prime },  \label{el2b}
\end{equation}%
where $r$ is defined implicitly by
\[
\tan v^{\prime \prime }\tan w^{\prime \prime }=-\exp \left[ \frac{%
r_{+}^{1}-r_{-}^{1}}{2(r_{+}^{1})^{2}}r\right] \sqrt{\frac{r-r_{+}^{1}}{%
(r-r_{-}^{1})^{\chi }}},\chi =\left( \frac{r_{+}^{1}}{r_{-}^{1}}\right) ^{2}.
\]%
As particular cases, we may chose $\eta _{3}\left( r,\varphi \right) $ as
the condition of vanishing of the metric coefficient before $\delta
v^{\prime \prime }\delta w^{\prime \prime }$ will describe a horizon
parametrized by a resolution ellipsoid hypersurface.

The maximal \ extension of the Schwarzschild metric deformed by a small
parameter $\varepsilon $ (for ellipsoid configurations treated as the
eccentricity), i. e. \ the extension of the metric (\ref{sch}), is defined
by taking (\ref{el2b}) as the metric on the maximal manifold on which this
metric is of smoothly class $C^{2}.$ The Penrose diagram of this static but
locally anisotropic space--time, for any fixed angular value $\varphi _{0}$
is similar to the Reissner--Nordstrom solution, for the case $%
e^{2}\rightarrow \varepsilon $ and $e^{2}<m^{2}$(see, for instance, Ref. %
\cite{haw})). There are an infinite number of asymptotically flat regions of
type I, connected by intermediate regions II and III, where there is still
an irremovable singularity at $r=0$ for every region III. We may travel from
a region I to another ones by passing through the 'wormholes' made by
anisotropic deformations (ellipsoid off--diagonality of metrics, or
anholonomy) like in the Reissner--Nordstrom universe because $\sqrt{%
\varepsilon }$ may model an effective electric charge. One can not turn back
in a such travel.

It should be noted that the metric (\ref{el2b}) $\ $\ is analytic every were
except at $r=r_{-}^{1}.$ We may eliminate this coordinate degeneration by
introducing another new coordinates%
\[
v^{\prime \prime \prime }=\arctan \left[ \exp \left( \frac{%
r_{+}^{1}-r_{-}^{1}}{2n_{0}(r_{+}^{1})^{2}}v\right) \right] ,w^{\prime
\prime \prime }=\arctan \left[ -\exp \left( \frac{-r_{+}^{1}+r_{-}^{1}}{%
2n_{0}(r_{+}^{1})^{2}}w\right) \right] ,
\]%
where the integer $n_{0}\geq (r_{+}^{1})^{2}/(r_{-}^{1})^{2}.$ In \ these
coordinates, the metric is analytic every were except at $r=r_{+}^{1}$ where
it is degenerate.$\,$\ This way the space--time manifold can be covered by
an analytic atlas by using coordinate carts defined by $(v^{\prime \prime
},w^{\prime \prime },\theta ,\varphi )$ and $(v^{\prime \prime \prime
},w^{\prime \prime \prime },\theta ,\varphi ).$ Finally we note that the
analytic extensions of the deformed metrics were performed with respect to
anholonomc frames which distinguish such constructions from those dealing
only with holonomic coordinates, like for the usual Reissner--N\"{o}rdstrom
and Kerr metrics.

\section{Geodesics on Static Ellipsoid Backgrounds}

In this section we analyze the geodesic congruence of the metric (\ref{sch1}%
) with the data (\ref{data}), for simplicity, being linear on $\varepsilon ,$%
by introducing the effective Lagrangian (for instance, like in Ref. \cite%
{chan})%
\begin{eqnarray}
2L &=&g_{\alpha \beta }\frac{\delta u^{\alpha }}{ds}\frac{\delta u^{\beta }}{%
ds}=-\left( 1-\frac{2m}{r}+\frac{\varepsilon }{r^{2}}\right) ^{-1}\left(
\frac{dr}{ds}\right) ^{2}-r^{2}q(r)\left( \frac{d\theta }{ds}\right) ^{2}
\label{lagrb} \\
&&-\eta _{3}(r,\varphi )r^{2}\sin ^{2}\theta \left( \frac{d\varphi }{ds}%
\right) ^{2}+\left( 1-\frac{2m}{r}+\frac{\varepsilon \Phi _{4}}{r^{2}}%
\right) \left( \frac{dt}{ds}+\varepsilon \widehat{n}_{1}\frac{dr}{ds}\right)
^{2},  \nonumber
\end{eqnarray}%
for $r=r(\xi ).$

The corresponding Euler--Lagrange equations,
\[
\frac{d}{ds}\frac{\partial L}{\partial \frac{\delta u^{\alpha }}{ds}}-\frac{%
\partial L}{\partial u^{\alpha }}=0
\]%
are%
\begin{eqnarray}
&&\frac{d}{ds}\left[ -r^{2}q(r)\frac{d\theta }{ds}\right] =-\eta
_{3}r^{2}\sin \theta \cos \theta \left( \frac{d\varphi }{ds}\right) ^{2},
\label{lag2b} \\
&&\frac{d}{ds}\left[ -\eta _{3}r^{2}\frac{d\varphi }{ds}\right] =-\eta
_{3}^{\ast }\frac{r^{2}}{2}\sin ^{2}\theta \left( \frac{d\varphi }{ds}%
\right) ^{2}+\frac{\varepsilon }{2}\left( 1-\frac{2m}{r}\right) \left[ \frac{%
\Phi _{4}^{\ast }}{r^{2}}\left( \frac{dt}{ds}\right) ^{2}+\widehat{n}%
_{1}^{\ast }\frac{dt}{ds}\frac{d\xi }{ds}\right]   \nonumber
\end{eqnarray}%
and%
\begin{equation}
\frac{d}{ds}\left[ (1-\frac{2m}{r}+\frac{\varepsilon \Phi _{4}}{r^{2}}%
)\left( \frac{dt}{ds}+\varepsilon \widehat{n}_{1}\frac{d\xi }{ds}\right) %
\right] =0,  \label{lag3}
\end{equation}%
where, for instance, $\Phi _{4}^{\ast }=\partial $ $\Phi _{4}/\partial
\varphi $ we have omitted the variations for $d\xi /ds$ which may be found
from (\ref{lagrb}). The sistem of equations (\ref{lagrb})--(\ref{lag3})
transform into the usual system of geodesic equations for the Schwarzschild
space--time if $\varepsilon \rightarrow 0$ and $q,\eta _{3}\rightarrow 1$
which can be solved exactly \cite{chan}. For nontrivial values of the
parameter $\varepsilon $ and polarization $\eta _{3}$ even to obtain some
decompositions of solutions on $\varepsilon $ for arbitrary $\eta _{3}$ and $%
n_{1[1,2]},$ see (\ref{auxf4}), is a cumbersome task. In spite of the fact
that with respect to anholonomic frames the metric (\ref{sch1}) has the
coefficients \ being very similar to the Reissner--Nordstom solution the
geodesic behaviour, in our anisotropic case, is more sophisticate because of
anholonomy and ''elongation'' of partial derivatives. For instance, the
equations (\ref{lag2b}) \ state that there is not any angular on $\varphi ,$
conservation law if $\eta _{3}^{\ast }\neq 0,$ even for $\varepsilon
\rightarrow 0$ (which holds both for the Schwarzschild and
Reissner--Nordstom metrics). One follows from the equation (\ref{lag3}) the
existence of an energy like integral of motion, $E=E_{0}+$ $\varepsilon
E_{1},$ with%
\begin{eqnarray*}
E_{0} &=&\left( 1-\frac{2m}{r}\right) \frac{dt}{ds} \\
E_{1} &=&\frac{\Phi _{4}}{r^{2}}\frac{dt}{ds}+\left( 1-\frac{2m}{r}\right)
\widehat{n}_{1}\frac{d\xi }{ds}.
\end{eqnarray*}

The introduced anisotropic deformations of congruences of Schwarzschild's
space--time geodesics mantain the known behaviour in the vecinity of the
horizon hypersurface defined by the condition of vanishing of the
coefficient $\left( 1-2m/r+\varepsilon \Phi _{4}/r^{2}\right) $ in (\ref%
{el2b}). The simplest way to prove this is to consider radial null geodesics
in the ''equatorial plane'', which satisfy the condition (\ref{lagrb}) with $%
\theta =\pi /2,d\theta /ds=0,d^{2}\theta /ds^{2}=0$ and $d\varphi /ds=0,$
from which follows that%
\[
\frac{dr}{dt}=\pm \left( 1-\frac{2m}{r}+\frac{\varepsilon _{0}}{r^{2}}%
\right) \left[ 1+\varepsilon \widehat{n}_{1}d\varphi \right] .
\]%
The integral of this equation, for every fixed value $\varphi =\varphi _{0}$
is
\[
t=\pm r^{\Vert }+\varepsilon \int \left[ \frac{\Phi _{4}(r,\varphi _{0})-1}{%
2\left( r^{2}-2mr\right) ^{2}}-\widehat{n}_{1}(r,\varphi _{0})\right] dr
\]%
where the coordinate $r^{\Vert }$ is defined in equation (\ref{r1}). In this
formula the term proportional to $\varepsilon $ can have non--singular
behaviour for a corresponding class of polarizations $\lambda _{4},$ see the
formulas (\ref{hor1}). Even the explicit form of the integral depends on the
type of polarizations $\eta _{3}(r,\varphi _{0})$ and $n_{1[1,2]}(r),$ which
results in some small deviations of the null--geodesics, we may conclude
that for an in--going null--ray the coordinate time $t$ increases from $%
-\infty $ to $+\infty $ as $r$ decreases from $+\infty $ to $r_{+}^{1},$
decreases from $+\infty $ to $-\infty $ as $r$ further decreases from $%
r_{+}^{1}$ to $r_{-}^{1},$ and increases again from $-\infty $ to a finite
limit as $r$ decreses from $r_{-}^{1}$ to zero. We have a similar behaviour
as for the Reissner--Nordstrom solution but with some additional anisotropic
contributions being proportional to $\varepsilon .$ Here we also note that
as $dt/ds$ tends to $+\infty $ for $r\rightarrow r_{+}^{1}+0$ and to $%
-\infty $ as $r_{-}+0,$ any radiation received from infinity appear to be
infinitely red--shifted at the crossing of the event horizon and infinitely
blue--shifted at the crossing of the Cauchy horizon.

The mentioned properties of null--geodesics allow us to conclude that the
metric (\ref{sch}) (equivalently, (\ref{sch1})) with the data (\ref{data})
and their maximal analytic extension (\ref{el2b}) really define a black hole
static solution which is obtained by anisotropic small deformations on $%
\varepsilon $ and renormalization by $\eta _{3}$ of the Schwarzchild
solution (for a corresponding type of deformations the horizon of such black
holes is defined by ellipsoid hypersurfaces). We call such objects as black
ellipsoids, or black rotoids. They exists in the framework of general
relativity as certain vacuum solutions of the Einstein equations defined by
static generic off--diagonal metrics and associated anholonomic frames. This
property disinguishes them from similar configurations of Reissner--Norstrom
type (which are static electrovacuum solutions of the Einstein--Maxwell
equations) and of Kerr type rotating configurations, with ellipsoid horizon,
also defined by off--diagonal vacuum metrics (here we emphasized that the
spherical coordinate system is associated to a holonomic frame which is a
trivial case of anholonomic bases).

\section{Conclusions}

We proved that there are such small, with nonlinear gravitational
polarization, static deformations of the Schwarschild black hole solution
(for instance, to some resolution ellipsoid like configurations) which
preserve the horizon and geodesic behaviour, but slightly deforme the
spherical constructions. This means that we may state such parameters of the
exact solutions of vacuum Einstein equations defined by off--diagonal
metrics with ellipsoid symmetry, constructed in Refs. \cite{v,v1,vth}, as
the solutions would define static ellipsoid black hole configurations.

The new class of static ellipsoidal black hole metrics posses a number of
similarities with the Reissner--Nordstrom metric:\ The parameter of
ellipsoidal deformation may be considered as an effective electromagnetic
charge induced by off--diagonal vacuum gravitational interactions. Effective
electromagnetic charges and Reissner--Nordstrom metrics, induced by
interactions in the bulk of extra dimension gravity, were considered
recently in brane gravity \cite{maartens}. In this paper (see also Refs. %
\cite{vth}) we proved that such Reissner--Nordstrom like ellipsoid black
hole configurations may be constructed even in the framework of vacuum
Einstein gravity if off--diagonal metrics and anholonomic frames are
introduced into consideration.

We emphasize that the static ellipsoid black holes posses spherical topology
and satisfy the principle of topological censorship \cite{haw1}. They are
also compatible with the black hole uniqueness theorems \cite{ut}; at
asymptotics, at least for a very small eccentricity, such metrics transforms
into the usual Schwarzschild one. We note that the stability of static
ellipsoid black holes can be proved similarly by considering small
perturbations of the spherical black holes \cite{velp}. (On the stability of
the Schwarzschild solution see details in Ref. \cite{chan}.)

It is interesting to compare the off--diagonal ellipsoidal metrics with
those describing the distorted diagonal black hole solutions (see the vacuum
case in Refs. \cite{ms} and an extension to the case of non--vanishing
electric fields \cite{fk}). In our ellipsoidal cases the distorsion of
spacetime is caused by some anisotropic off--diagonal terms being
non--trivial in some regions but in the case of ''pure diagonal''
distorsions one treats such effects following the fact that the vacuum
Einstein equations are not satisfied in some regions because of presence of
matter. As we emphasized in the introduction section, the off--diagonal
gravity may model some gravity--matter like interactions (like in
Kaluza--Klein theory, for some very particular configurations and
topological compactifications) but, in general, the off--diagonal vacuum
gravitational dynamics can not be associated to any effective matter
dynamics in a holonomic gravitational background. So, we may consider that
the anholonomic ellipsoidal deformations of the Schwarzschild metric are
some kind of anisotropic off--diagonal distorsions modeled by certain vacuum
gravitational fields with the distorsion parameteres (equivalently, vacuum
gravitational polarizations) depending both on radial and angular
coordinates. In general, both classes of off--diagonal anisotropic and
''pure'' diagonal distorsions (like in Refs. \cite{ms}) result in solutions
which are not asymptotically flat. However, it is possible to find
asymptotically flat extensions, as it was shown in this paper, even for
ellipsoidal configurations by introducing the corresponding off--diagonal
terms (the asymptotic conditions for the diagonal distorsions are discussed
in Ref. \cite{fk}. To satisfy such conditions one has to include some
additional matter fields in the exterior portion of spacetime.)

Using the methods elaborated and developed in Refs. \cite{v,vth,vsbd}, and
in this paper, we can construct off--diagonal ellipsoidal extensions of the
already diagonally disturbed Schwarzschild metric (see the metric (3.7) from
Ref.\cite{fk}). Such anholonomic deformations would contain in the diagonal
limit configurations with $\varepsilon \rightarrow 0$ but $\eta _{3}\neq 1;$
for such configurations the function $\eta _{3}$ has to be related in the
corresponding limits with the values $\overline{\gamma }_{D},\overline{\psi }%
_{D}$ and $A$ from \cite{fk}. We remark that there are different classes of
ellipsoidal deformations of the Schwarschild metric which result in a vacuum
configuration. The conditions $\varepsilon \rightarrow 0$ and $q,\eta _{3}=1$
select just the limit of the usual radial Schwarschild asymptotics without
any (also possible) additional diagonal distorsions.

The deformation parameter $\varepsilon $ effectively seems to put an
''electric charge'' on the black hole which is of gravitational
off--diagonal/anholonomic origin. It can describe effectively both positive
and negative gravitational polarizations (even some repulsive gravitational
effects). This is not surprising because the coefficients of an anisotropic
black hole are similar to those of the Reissner--Nordstrom solution only
with respect to corresponding anholonomic frames which are subjected to some
constraints (anholonomy conditions). Intuitively, we may compare such
effects with those from the usual Newton gravity: a ball falls directly on
the Earth but it can run under an angle inside a tube because of constraints
imposed at the boundary (by anholonomic frames in general relativity we may
model a more sophisticate behaviour with locally anisotropic gravitational
polarizations and even repulsion).

For the ellipsoidal metrics with the Schwarzschild asymptotics, the
ellipsoidal character could result in some observational effects in the
vicinity of the horizon (for instance, scattering of particles on a static
ellipsoid; we can compute anisotropic matter accretion effects on an
ellipsoidal black hole put in the center of a galactic being of ellipsoidal
or another configuration). A point of further investigations could be the
anisotropic ellipsoidal collapse when both the matter and spacetime are of
ellipsoidal generic off--diagonal symmetry (former theoretical and
computational investigations were performed only for rotoids with
anisotropic matter and particular classes of perturbations of the
Schwarzshild solutions \cite{st}). For very small eccentricities, we may not
have any observable effects like perihelion shift or light bending if we
restrict our investigations only to the Schwarzshild--Newton asymptotics.

Finally, we present some comments on mechanics and thermodynamics of
ellipsoidal black holes. For the static black ellipsoids with flat
asymptotics, we can compute the area of the ellipsoidal horizon, associate
an entropy and develop a corresponding black ellipsoid thermodynamics. But
this would be a very rough approximation because, in general, we are dealing
with off--diagonal metrics depending anisotropically on two/three
coordinates. Such solutions are with anholonomically deformed Killing
horizons and should be described by a thermodynamics (in general, both
non-equilibrium and irreversible) of black ellipsoids self--consistently
embedded into an off--diagonal anisotropic gravitational vacuum. This is a
ground for numerous new conceptual issues to be developed and related to
anisotropic black holes and the anisotropic kinetics and thermodynamics \cite%
{v1} as well to a framework of isolated anisotropic horizons \cite{asht}
which is a matter of our further investigations.


\subsection*{Acknowledgements}

~~ The work is supported by a NATO/Portugal fellowship at CENTRA, Instituto
Superior Tecnico, Lisbon. The author is very grateful to the referee who
pointed to very important references and subjects for research and
discussion.



\end{document}